\documentclass[iop,apj]{emulateapj}
 \usepackage{natbib}
\bibpunct{(}{)}{;}{a}{}{,} 

\slugcomment{Submitted to the Astrophysical Journal}
\shorttitle{Non-thermal emission in a planetary nebula}
\shortauthors{Su\'arez et al.}

\begin{document}

\title{Time-variable non-thermal emission in the planetary nebula IRAS
  15103$-$5754}

\author{Olga Su\'arez\altaffilmark{1}, Jos\'e F. G\'omez\altaffilmark{2}, 
Philippe Bendjoya\altaffilmark{1}, Luis F. Miranda\altaffilmark{2,3}, 
Mart\'{\i}n A.  Guerrero\altaffilmark{2}, Lucero Uscanga\altaffilmark{4},
James A.  Green\altaffilmark{5,6}, J. Ricardo Rizzo\altaffilmark{7}, 
Gerardo Ramos-Larios\altaffilmark{8}}

\altaffiltext{1}{Laboratoire Lagrange, UMR 7293, Universit\'e de Nice Sophia-Antipolis, CNRS, Observatoire de la C\^ote d'Azur, 06304, Nice, France}
\altaffiltext{2}{Instituto de Astrof\'{\i}sica de Andaluc\'{\i}a (IAA-CSIC), Glorieta de la Astronom\'{\i}a S/N, 18008, Granada, Spain}
\altaffiltext{3}{Universidad de Vigo, Departamento de F\'{\i}sica aplicada,
   Facultad de Ciencias, Campus Lagoas-Marcosende s/n, 36310, Vigo,
   Spain}
\altaffiltext{4}{Institute of Astronomy, Astrophysics, Space Applications and Remote Sensing, National Observatory of Athens, 15236 Athens, Greece}
\altaffiltext{5}{CSIRO Astronomy and Space
Science, Australia Telescope National Facility, PO Box 76, Epping, NSW
1710, Australia}
\altaffiltext{6}{SKA Organisation, Jodrell Bank Observatory, Lower Withington, Macclesfield SK11 9DL, UK}
\altaffiltext{7}{Centro de Astrobiolog\'{\i}a (INTA-CSIC),
Ctra. M-108, km.~4, E-28850 Torrej\'on de Ardoz, Spain}
\altaffiltext{8}{Instituto de Astronom\'{\i}a y Meteorolog\'{\i}a, Av. Vallarta No. 2602, Col. Arcos 
Vallarta, C.P. 44130 Guadalajara, Jalisco, Mexico}

\begin{abstract}
  The beginning of photoionization marks the transition between the
  post-Asymptotic Giant Branch (post-AGB) and planetary nebula (PN)
  phases of stars with masses $\la 8$ M$_\odot$.  This critical phase
  is difficult to observe, as it lasts only a few
  decades.  The combination of jets and magnetic fields, the key
  agents of PNe shaping, could give rise to
  synchrotron emission, but this has never been observed before in any
  PNe, since free-free emission from the ionized gas is expected to
  dominate its radio spectrum.  In this paper we report radio
  continuum observations taken with the Australia Telescope Compact
  Array between 1 and 46 GHz of the young PN IRAS 15103$-$5754.  Our
  observations in 2010-2011 show non-thermal emission compatible with
  synchrotron emission from electrons accelerated at a shock with spectral
  index $\alpha\simeq -0.54$. However, in 2012, the spectral index $\alpha \simeq
  -0.28$ is no longer compatible with synchrotron emission in these
  types of processes.  Several hypothesis are discussed to explain
  this change. The more plausible ones are related to the presence of
  the newly photoionized region in this young PN: either energy loss
  of electrons due to Coulomb collisions with the plasma, or selective
  suppression of synchrotron radiation due to the Razin effect. We
  postulate that the observed flattening of non-thermal radio spectra
  could be a hallmark identifying the beginning of the PN phase.


\end{abstract}

\section{Introduction}

Planetary Nebulae (PNe) represent one of the final phases in the
evolution of stars with masses $\la 8$ M$_\odot$, when they become
hot enough to photoionize the circumstellar envelope ejected during
the Asymptotic Giant Branch (AGB).  This nebular photoionization
proceeds very quickly, over only a few decades \citep{bobrowsky98},
making it extremely difficult to study the events produced just at
the transition to the PN phase, which are crucial to determine how PNe
form.
  
Many PNe exhibit complex axisymmetric or multipolar morphologies that
are believed to be caused by collimated jets \citep{sahai98} launched
during the post-AGB (which just precedes the photoionization of the
envelope) and PN phases. The exact mechanism for the launching of
these jets is still unknown, but possible scenarios are the presence
of magnetic fields \citep{garciasegura05} and/or binary systems
\citep{soker00,nordhaus06,demarco09}.  The combination of jets and
magnetic fields lead to the presence of synchrotron radiation in
several astrophysical contexts, such as active galactic nuclei
\citep{bridle84}, microquasars \citep{mirabel92}, or massive
protostars \citep{carrascogonzalez10}.  This emission is caused by the
motion within a magnetic field of electrons accelerated
to relativistic velocities at shock fronts induced by jets
\citep{achterberg00}.

Magnetic fields have been detected in the circumstellar
envelopes of some PNe
\citep{miranda01,sabin07}, 
but synchrotron emission has never been unambiguously reported in PNe
before \citep[see, e.g.,][for a review of previous attempts for such a
detection]{casassus07}, even though the energy conditions in PNe are
well-suited for the production of synchrotron radiation
\citep{dgani98,casassus07}. The reason is that the strong free-free radiation from
ionized gas that characterize these objects is expected to overwhelm
any type of non-thermal emission. \cite{casassus07} claimed that
synchrotron emission is a possible source of centimeter-wave excess in
the radio continuum flux of some PNe, over what would be expected from
free-free emission alone, but the available data could not confirm
this hypothesis.  On the other hand, non-thermal emission has been
observed in a few post-AGB sources \citep{cohen06,bains09,
  perezsanchez13_synchrotron}, where photoionization has not yet
started but jets are already present. All this indicates that the
temporal window for a possible study of synchrotron radiation in PNe
is extremely short, since it cannot be detected very shortly
after the onset of photoionization.

The source IRAS 15103$-$5457 (hereafter I15103) is an optically
obscured PN showing a high-velocity jet traced by water masers
\citep{bendjoya14,gomez15}.  Water maser jets are seen in
a few late AGB and post-AGB stars, and are termed ``water fountains''
\citep{imai07}. I15103 is the first known case of a ``water fountain''
that has already entered the PN phase, suggesting that it may be one
of the youngest PN known. Therefore, it is an excellent candidate to
study the processes taking place during the transition into this
phase. Moreover, interferometric radio continuum observations obtained
in the Red MSX Survey \citep[RMS,][]{lumsden13} show a flux
density lower at 3.6 cm than at 6 cm \citep{urquhart07}, which suggests the
emission may be non-thermal. A confirmation of this trend was needed,  
since only two wavelengths were observed in this
survey, and problems such as calibration uncertainties or extended flux
missed by the interferometer may have affected these values of flux density.

In order to confirm the presence of non thermal emission in I15103, we
present new radio continuum observations covering a much wider range
of wavelengths in different epochs. In Section 2 we describe the
observations performed. In Section 3 we explain the results derived
from the observations. Section 4 is devoted to the different
hypothesis we analyse to explain the variations observed in the
spectral energy distribution (SED). These hypotheses are put in
context and the predictions of the future evolution of the SED are
presented. Section 5, finally, presents the conclusions of this work.

\section{Observations and data reduction}

Radio observations towards I15103 were performed using the
Australia Telescope Compact Array (ATCA) in several epochs between
2010 and 2012. Table \ref{table_observations} summarizes the different
frequencies and array configuration used. All observations used the
Compact Array Broadband Backend (CABB), which provides two independent
intermediate frequency (IF) outputs of 2 GHz in dual linear
polarization. In the particular case of observations around
2.1 GHz, both IF were centered at the same frequency, since each IF
covers the whole frequency band of the receiver (to
provide redundancy in case of correlator failure), so we only processed
one of them. The observations on August 2011 used the 64M mode of
CABB, which samples each 2 GHz bandpass into 32 frequency channels of
64 MHz. All other observations used the 1M mode, which samples 2048
channels, each of them 1 MHz wide. %

\begin{deluxetable}{lcc}
\tablecolumns{3}
\tablewidth{0pc}
\tablecaption{Observations performed \label{table_observations}}
\tablehead{
\colhead{Date}    &  \colhead{Frequency\tablenotemark{a} }   & \colhead{Array\tablenotemark{b}}\\
                  & \colhead{(GHz)}         & \\
}
\startdata
2010-Oct-07  & 5.5, 9.0    & H214  \\
                  &43.0, 45.0 & H214\\
2011-Aug-11 & 21.8, 24.0 & H168 \\
2012-Oct-23 & 2.1          & H214 \\
                & 5.5, 9.0     & H214\\
                   & 22.0, 24.0   & H214\\
2012-Dec-10   & 2.1       & 6B \\
                       & 5.5, 9.0       & 6B \\
                &   26.0  & 6B \\  
 \enddata
 \tablenotetext{a}{Central frequency of each band.}
  \tablenotetext{b}{ATCA configuration used.}
\end{deluxetable}

The complex gain calibrator was PKS 1613-586 in all datasets. The
absolute flux calibrator was PKS 1934-638 at all observed frequencies,
except at the beginning of the observations centered at 5.5 and 9.0 GHz
on December 2012, when this source was not yet above the horizon, and
PKS 0823-500 was used instead for this purpose. Since PKS 1934-638 is
the recommended source to set the flux density scale in ATCA data, in
these particular set of data, we independently calibrated the data
that used PKS 1934-638 and PKS 0823-500 as flux calibrators, and
rescaled the latter data (by a factor of $\simeq 1.10-1.15$, depending
on frequency) to achieve the same final flux density on the complex gain
calibrator throughout the observations that day. This indicates that
the flux density model of PKS 0823-500 used by the data reduction
package was in error by $\simeq 10-15$\%, as pointed out in the ATCA
calibrator catalog. The two sets of uv-data were merged after this
rescaling. The flux calibrators were also used to calibrate the
bandpass at frequencies below 15 GHz. At higher frequencies, we
initially used PKS 1921-293 to correct the bandpass, but checked the
results against PKS 1934-638, to confirm that the resulting spectral
slopes in our data were correct.

All data were calibrated and processed with the MIRIAD package using
standard procedures. Calibration, initial imaging and self-calibration
were performed across the entire 2 GHz bands. Final uv-data were then
imaged in successive 200-MHz intervals to obtain the intra-band
variation of flux density with frequency. The two exceptions for this
processing mode were the data between
43$-$45\,GHz, where the signal-to-noise ratio was lower and we
present the average of each 2 GHz band, and the data between 21-25 GHz
taken on August 2011 where we kept the original 64 MHz spacing
provided by the correlator. To double check our results, we
reprocessed the data by splitting the raw uv-data into the same
frequency intervals, which were then independently calibrated,
self-calibrated, and imaged. The resulting dependence of flux density
with frequency was the same using both methods of splitting. In all
cases, broadband data were weighted with robust parameter 0.5 and
frequency synthesis was applied, using MIRIAD task ``invert''. Images
were subsequently deconvolved with the CLEAN algorithm, as implemented
in task ``mfclean''.

Low-frequency observations are contaminated by extended Galactic
background emission. This affects the shortest baselines. In order to
obtain reliable maps we had to restrict the data to baselines $>6-8$
k$\lambda$ at frequencies $<3.5$ GHz. With this
  restriction, the 2.1 GHz data taken in October 2012 with the compact
  H214 array were unusable. We also note that a visual inspection of
the maps between 4.5 and 6.5 GHz obtained on October 2010, revealed
significant artifacts and a higher noise (a factor of 3) than those at
$>8$ GHz. We think that they could be affected by low-level
radio-frequency interferences (RFI), which we could not identify in
uv-data for flagging, but that affects the data
reliability. Furthermore, the corresponding flux density points show a
relatively large scatter.

To reconstruct the SED behavior with
time we have used the archival and published data summarized in Table
\ref{table_archive}. These data have been taken with ATCA by other
authors in 1991 \citep{vdsteene93}, 1999 \citep{gaensler00}, 2002 and
2004 \citep{urquhart07}, as well as with the Molonglo Observatory
Synthesis Telescope in 1998 for the Second Molonglo Galactic Plane
Survey (1998) \citep{murphy07}. We used the published flux densities
for the data from 1991 and 1998. We note that data from 1991 do not have
enough uv coverage to obtain images, so their flux density values may
have a significantly larger error than quoted. We downloaded and
processed the ATCA archival data on 1999 and 2002 to obtain the values
in Table \ref{table_archive}. The fluxes on 2004 were measured on
processed images provided by the authors \citep{urquhart07}.  The data
taken in January 2012 centered at 1.8 GHz for the MAGMO project
\citep{green12}, included I15103 within their primary beam, with their
phase center located $13.7^{\prime}$ away from our source. Therefore,
we processed these data correcting by the response of the ATCA primary
beam as implemented in the MIRIAD task ``linmos''.

\begin{deluxetable}{lcccc}
\tablecolumns{5}
\tablewidth{0pc}
\tablecaption{Archival data \label{table_archive}}
\tablehead{
\colhead{Date}    &  \colhead{Frequency\tablenotemark{a}}   & \colhead{Flux density} & \colhead{rms\tablenotemark{b}} & References\\
                  & \colhead{(GHz)}                         & (Jy)                   & (Jy)\\
}
\startdata
1991.4        & 4.8 & 0.1145 & 0.0008   & 1\\
1998.4        & 0.8 &0.193  & 0.006  & 2\\
1999.2        &1.4  & 0.199 & 0.003  & 3 \\
                & 2.5 & 0.219 & 0.004 & 3\\
2002.7       & 4.8 &0.1509 & 0.0006 & 4 \\
                &  8.4 &0.1088 & 0.0004  & 4\\  
2004.9       & 4.8 & 0.0989 & 0.0007& 4 \\
                &  8.4 & 0.068 & 0.002& 4\\  
2012.0        &  1.8   & 0.052\tablenotemark{c} & 0.002\tablenotemark{c} & 5\\

 \enddata
 \tablenotetext{a}{Central frequency of the observation}
  \tablenotetext{b}{One-sigma rms noise of the map}
   \tablenotetext{c}{Mean value for the frequency intervals in which we split the whole band}
  \tablerefs{1. \cite{vdsteene93}. 2. \cite{murphy07}. 3. \cite{gaensler00} 4. \cite{urquhart07}. 5. \cite{green12}}
\end{deluxetable}

\section{Results}

\begin{figure*}
\plotone{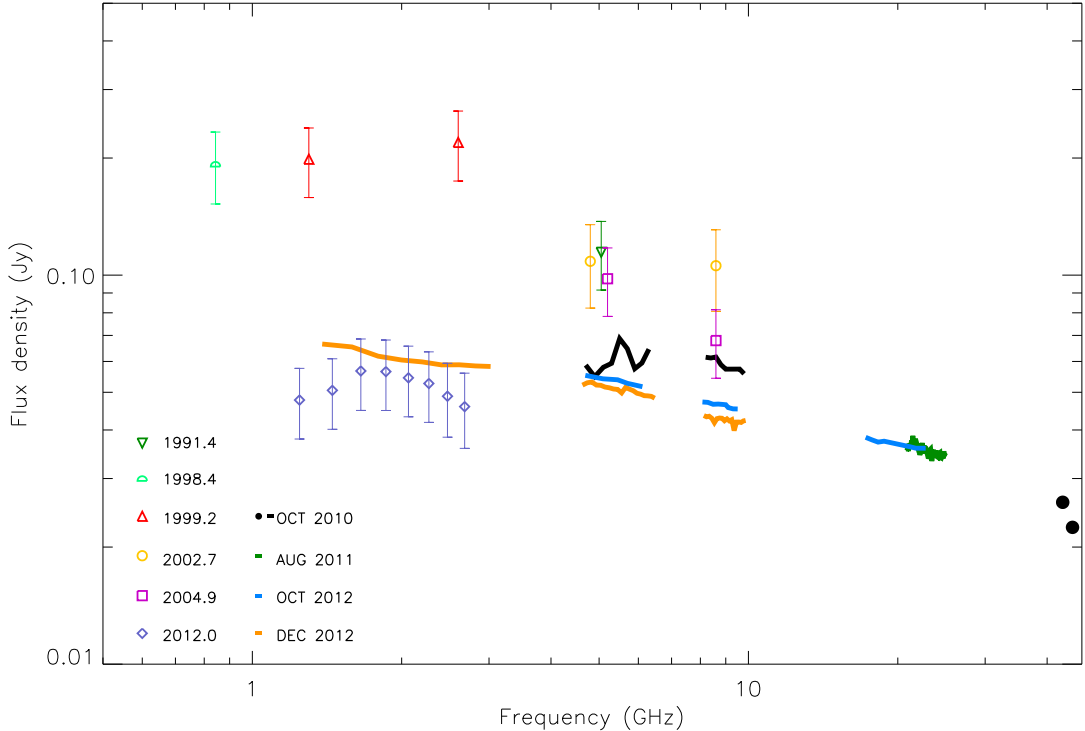}
\caption{Spectral energy distribution and flux density time-variation
  of I15103 at radio wavelengths. Our data are represented by solid
  lines: in black the October 2010 data (including filled circles for
  data at 43-45 GHz); green, August 2011; blue, October 2012 and orange,
  December 2012. The symbols represent archival and literature data as
  listed in Table 2. Flux density uncertainties in our data are
  $\simeq 10\%$. The three points at 4.8 GHz
  from archives and literature have been slightly displaced in frequency in this figure for clarity, but they were taken at the same frequency. \label{fig_observations}}
\end{figure*}

In Figure \ref{fig_observations} we present the radio continuum emission
in I15103 at different frequencies and epochs, obtained with our
observations and the archival data.  Our radio continuum observations
show a dependency of flux density $S_\nu$ as a function of frequency
that is atypical for a PN. Radio continuum emission from PNe normally
consists in free-free (bremsstrahlung) radiation in the ionized gas,
and follows a relation with frequency $S_\nu\propto \nu^\alpha$, where
the spectral index $\alpha$ ranges from $\alpha\simeq 2$ in the
optically thick regime at low frequencies (usually $\nu \la$ 10~GHz),
to $-0.1$ for optically thin emission at high frequencies (usually
$\nu \ga$ 10~GHz).  The turnover frequency separating these two
opacity regimes is thought to depend on the age of the PN
\citep{kwok81}.

However, our data taken in the first two sets of observations
(2010-2011; black and green solid lines and black filled
circles) show radio continuum emission with a steep decline at high
frequencies (Figure \ref{fig_observations}). We found a spectral index
$\alpha=-0.54\pm 0.08$ between 8 and 25 GHz, indicating a non-thermal
origin for the emission.  The data at higher frequencies (42-46 GHz)
show an apparently steeper spectral index, but given that it is
obtained from only two points, its significance is low ($\alpha\simeq
-3\pm3$).  At lower
frequencies (4-5.5 GHz), the data have lower quality, with artifacts
in the maps that may indicate the presence of low-level RFI, as
explained above, so the possible turnover suggested by the data at
these frequencies is not reliable.

Our most recent ATCA data (October-December 2012), taken about one
year later, show a significantly flatter spectrum over the whole range
from 1 to 24 GHz. In particular, the data taken on October 2012 (blue
solid lines in Fig. \ref{fig_observations}) follow a spectral
index of $\alpha=-0.28\pm 0.08$ between 4.5 and 25 GHz.  The slope is
similar for data taken on December 2012 between 1 and 10 GHz (orange solid
  lines in Fig. \ref{fig_observations}). 

\begin{figure}
\epsscale{0.9}
\plotone{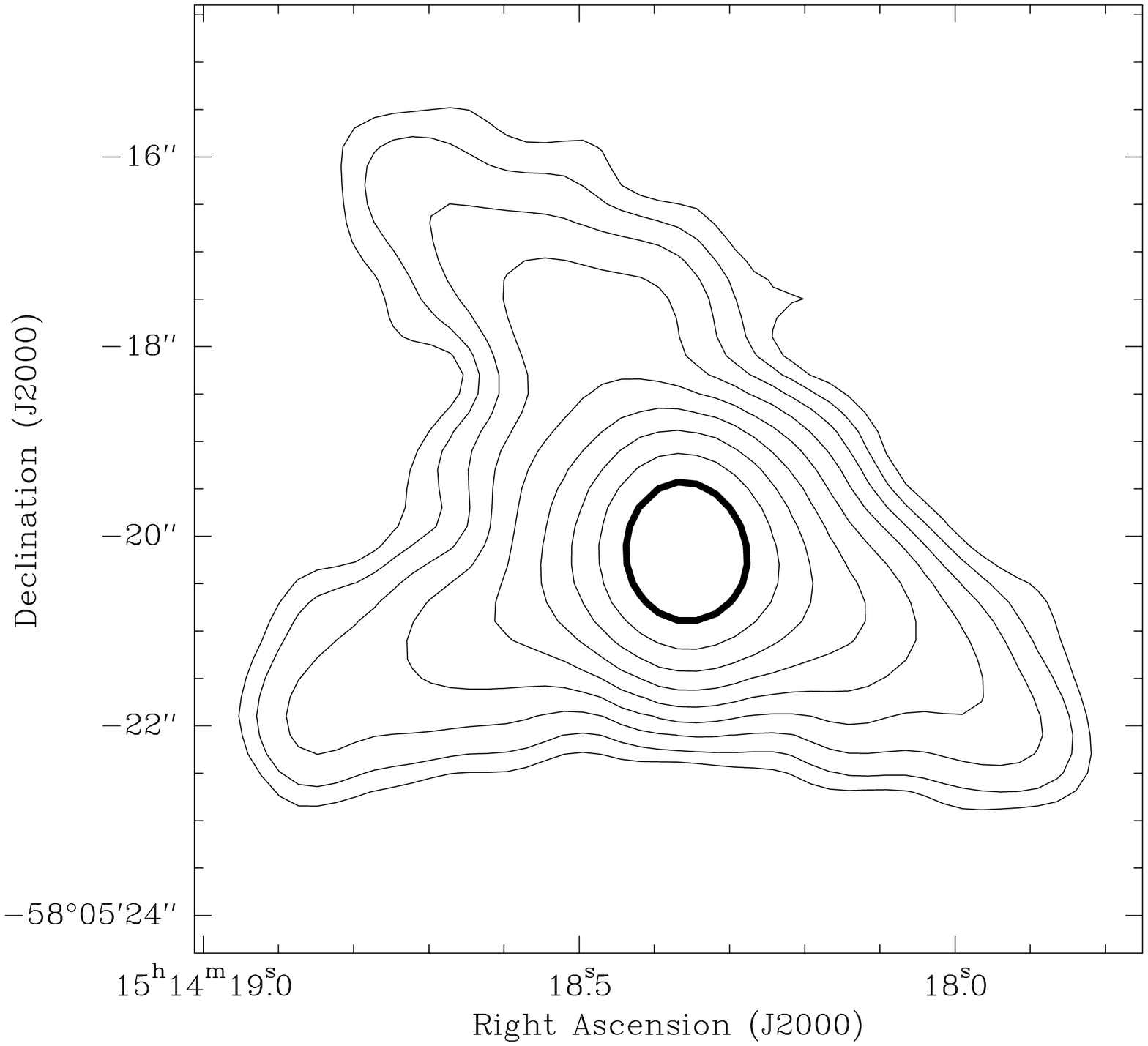}
\epsscale{1.0}
\plotone{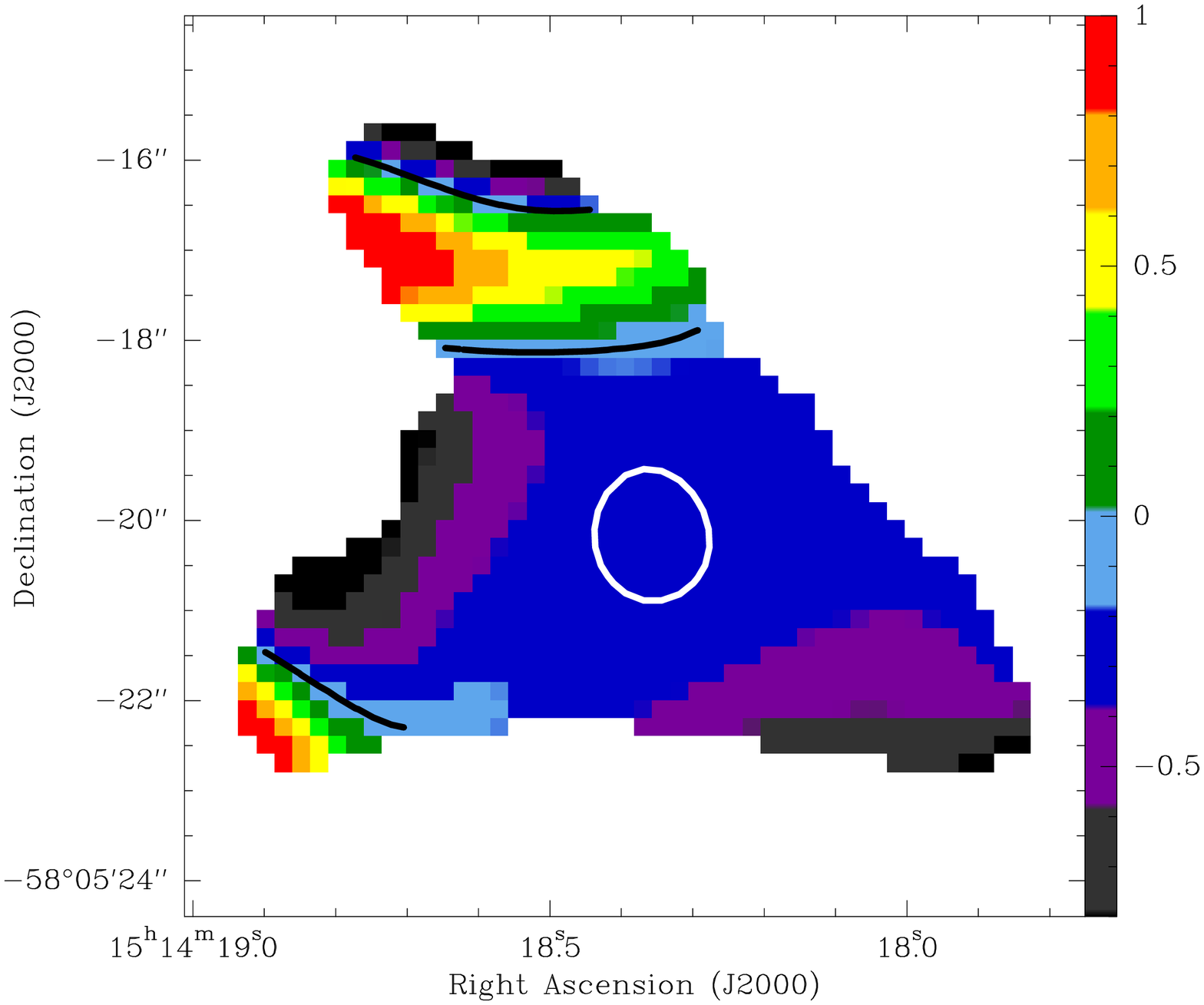}
\caption{Top: Map of radio continuum emission integrated between 8 and
  10 GHz. Contour levels are $2^n$ (with $n$ from 0 to 8) times the
  3$\sigma$ noise level ($\sigma=1.8\times 10^{-5}$ Jy
  beam$^{-1}$). The beam size is $1.4''\times 1.1''$, (position angle =
  $-3^\circ$). The thick contour ($1.38\times 10^{-2}$ Jy beam$^{-1}$)
  nearly corresponds to the half-power level. The maximum of the image
  is $2.73\times 10^{-2}$ Jy beam$^{-1}$. Bottom: Spectral index map
  obtained with the radio continuum emission at 8-10 GHz and at
  4.5-6.5 GHz, both convolved to a common angular resolution of
  $2.25''$. Black contours correspond to the spectral indices of $-$0.1,
  delimiting the areas that may correspond to thermal emission. The
  white contour is the half-power level of the radio continuum map in
  the top panel. \label{fig_map} }
\end{figure}

  The data taken on December 2012 allow us to resolve some structure
  in the radio continuum emission. Figure \ref{fig_map} (top) shows a
  map of the emission integrated between 8 and 10 GHz. The emission is
  dominated by an unresolved intense core, but there is much weaker
  emission extending along three arms toward the northeast, southwest,
  and southeast. Note that the contour levels in this map are not
  linear, but proportional to powers of 2, and the highest level
  nearly corresponds to the half power level. Figure \ref{fig_map}
  (bottom) shows a spectral index map obtained with the emission
  between 8 and 10 GHz (map in Fig. \ref{fig_map}-top), and the radio
  continuum emission between 4.5 and 6.5 GHz, both convolved to a
  common angular resolution of $2.25''$. As expected in view of the
  SED, most of the emission is non-thermal (with $\alpha < -0.1$),
  including the central bright core. However, there are some areas
  that are compatible with thermal emission ($\alpha > -0.1$),
  specially toward the northeast (the positive spectral indices toward
  the southeast are at lower intensity levels and are less reliable).
  The intensity of this possible thermal emission is more than 50
  times weaker than the non-thermal peak, and it does not
  significantly affect the derived spectral index from
  Fig. \ref{fig_observations}.

The time evolution of flux density derived from archival observations
is also consistent with the emission becoming weaker with time, a
variation that seems to be dramatic at low frequencies. The only
exception to this historical weakening is the data from the MAGMO
project, taken in 2012, and optimized for spectral line work.  The
MAGMO data suggest a turnover, and a slightly lower flux density than
our observations at these frequencies, which were taken
later. However, we consider these flux densities less reliable, given
the relatively large distance of the source from the phase center of
the observations, and the use of a different primary flux calibrator: PKS 0823-500 instead of PKS 1934-638 in our data. As noted above,
the flux density model of PKS 0823-500 in MIRIAD might be incorrect, and so the combination of positional offset
and calibrator flux density error margin reduce the significance of
any differences. 

Apart from the flux density scale, the flux density of the MAGMO
  data also shows an apparent turnover around 1.7 GHz. An incorrect
  spectral index in the model of PKS 0823-500 could produce such an
  effect. On the other hand, the turnover could be real, but it would
  have to be a very short-lived phenomenon, since we did not see it in
  our observations in December 2012, not even a year after the MAGMO
  ones. The unconfirmed result makes it difficult to ascertain
  the reliability of the possible turnover.

In any case, the global flattening of the flux density as a function of
frequency, which we see in our observations in a timescale of 1-2
years (Fig.\ \ref{fig_observations}) is a robust result, since the
same slope is seen in three different frequency ranges, observed
independently.

\section{Discussion}

Several scenarios are possible for the origin and variations of the
thermal and non-thermal radio continuum emission detected at
I15103. We describe in the following the different hypothesis and
their implication in the global context of stellar evolution.

\subsection{Thermal emission from ionized gas}

The weak emission detected on the radio continuum map with spectral
indices $\alpha> -0.1$ (Fig.~\ref{fig_map}) is compatible with the
typical free-free emission from ionized gas in PNe. It could
represent photoionized gas extending toward the northeast lobe of the
mid-IR nebulosity \citep{lagadec11}, in the direction of the water maser jet
\citep{gomez15}. The overall emission in the source, however, is
strongly dominated by non-thermal emission.

\subsection{The origin of non-thermal emission and its variation}
\label{sec_contribution}

A very striking characteristic of the
radio emission in I15103 is that it mostly shows negative spectral
indices (see Fig.~\ref{fig_map}), which indicates that it is dominated by non-thermal
processes.  This emission should arise from electrons moving in a
magnetized medium.  In particular, the spectral index $\alpha=-0.54\pm
0.08$ found between 8 and 25 GHz in 2010-2011 is consistent with
either synchrotron emission from relativistic electrons, or
gyrosynchrotron from mildly relativistic ones. 
The spectral index
both in the case of synchrotron or gyrosynchrotron emission depends on
the power-law energy distribution of electrons ($N(E)\propto
E^{-\beta}$). Therefore, a change in spectral index, as observed
between 2010 and 2012, could indicate a
time variation in this energy distribution that can be used to discern
which processes are at work. Alternatively, it could be
the result of absorption or suppression processes that are stronger on
radiation at lower frequencies. We will discuss these possibilities in
the following.

\subsubsection{Synchrotron emission and Coulomb collisions}
\label{sec_coulomb}
The spectral index $\alpha=-0.54\pm 0.08$ is
similar to that found in synchrotron emission in galaxies with active
nuclei \citep{bridle84}. In the case of pure synchrotron emission, the
spectral index can be related to the energy distribution of
electrons, in the form
$\alpha=(1-\beta)/2$. In our case, the observed $\alpha$ would imply a
value of $\beta = 2.08$.  

In the energy conditions of PNe, mass-loss processes can produce
collisionless, non-relativistic shocks and electrons may undergo
diffusive shock acceleration \citep{drury83,achterberg00} up to very
high velocities (this process is also called ``first-order Fermi
acceleration''). In this case, the exponent of the energy distribution
of electrons is given by
\begin{equation}
\beta=\frac{r+2}{r-1}
\end{equation}
where $r$ is the compression ratio of the gas at the shock. The
maximum value of $r$ in an ideal, non-relativistic gas is $\simeq
4$. The value $\beta=2.08$ we obtained indicates that $r$ is close to
this maximum value.

In the case of the data taken on October 2012, the spectral index of
$\alpha = -0.28\pm 0.08$ found between 4.5 and 25 GHz would imply an
index for the energy distribution of electrons, $\beta \simeq 1.56$ if
the emission were due to synchrotron emission. However, since the
maximum compression ratio in collisionless shocks is $r=4$, diffusive
shock acceleration cannot explain indices $\beta<2$.  Therefore, the
spectral index in the pure synchrotron regime should always be
$\alpha<-0.5$, while the acceleration is taking place.

However, the synchrotron emission could be the result of a shock that
took place during a short time and electrons are not being accelerated
any longer. In that case, the observed spectral flattening reflects
the evolution of electron energy after the shock ended. We note that
I15103 is a PN \citep{gomez15}, and it has developed a photoionized region around
it. When the relativistic electrons move in a dense plasma, they lose
energy due to Coulomb collisions against ions. The lifetime of
electrons of lower energy is shorter under Coulomb losses
\citep[e.g.,][]{rephaeli79,sarazin99}:
\begin{equation}
t_{\rm cou}(\gamma ) \simeq \gamma \left(1.2\times 10^{-12} n_e \left[ 1.0+ \frac{\ln (\gamma/n_e)}{75} \right]\right)^{-1}
{\rm s}, \label{eq_timescale}
\end{equation}
were $n_e$ is the electron density of the plasma expressed in
cm$^{-3}$, and $\gamma$ is the Lorentz factor of the
relativistic electrons. The result is a progressive hardening of the
energy distribution, with less electrons of low energy. This will lead
to a flattening of the synchrotron spectrum. Spectral flattening due
to Coulomb collisions is also observed in the radio spectrum of solar
flares \citep{melnikov98,lee00}.

The lifetimes of electrons under Coulomb collisions are consistent
with the observed flattening of I15013 over time. In particular,
assuming a magnetic field of $\simeq 5$ mG, as seen in some post-AGB stars
\citep{bains03, bains04}, according to  \citet{shu91}, the Larmor frequency of electrons
\begin{equation}
\nu_L = \frac{e B}{2\pi m_e c}
\end{equation}
(where $e$ and $m_e$ are the electron charge and mass, respectively,
and $B$ is the magnetic field strength) would be $\simeq 14$
kHz. Thus, the synchrotron emission at $\simeq 1$ GHz would typically
be due to electrons with $\gamma = \sqrt{\nu/\nu_L} = 267$.

Equation \ref{eq_timescale} indicates that the lifetime of these
electrons under Coulomb collisions and assuming electron density of
$2\times 10^5$ cm$^ {-3}$ \citep[e.g.][]{tafoya09} would be $\simeq 40$
yr. This is a reasonable value considering the timescales of
variations at low frequency shown in
Fig. \ref{fig_observations}. Under this interpretation, the observed
non-thermal emission in I15103 would be due to synchrotron emission from
electrons accelerated during an explosive event that occurred a few years
ago, and the emission would now be decaying due to Coulomb losses. 

We note that a recent explosive event would also explain the velocity
gradient of the maser emission in this source \citep{gomez15}, which
reinforces this interpretation. 

We also note that the timescale of relativistic electrons will shorten
with increasing plasma density (equation \ref{eq_timescale}). The
change in spectral index between 2010 and 2012 seems particularly sudden
compared with the evolution in previous years. This is suggestive of
an increasing density of ionized gas, consistent with I15103 being a
very young PN that is developing its photoionized region.

\subsubsection{Gyrosynchrotron emission from mildly relativistic electrons}

An alternative explanation to the observed change in the SED of I15103
would be that we witness the variation in gyrosynchrotron emission from
mildly relativistic electrons. This radiative process is seen in
active stars \citep[e.g.,][]{dulk85,gunn94,pestalozzi00}. The flux density
of these stars roughly follows a power-law of the form
\begin{equation}
S_\nu \propto 10^{-0.52\beta}(\sin\theta)^{-0.43+0.65\beta}\left[\frac{\nu}{\nu_b}\right]^\alpha,
\label{eq_gyro}
\end{equation}
where the relation between the spectral index and the distribution of
electrons is given by $\alpha \simeq 1.22-0.90\beta$ \citep{dulk85},
$\theta$ is the pitch angle between the electron trajectory and the
magnetic field, and $\nu_b\simeq 2.8\times 10^6 B$ is the
electron-cyclotron frequency.

Equation (4) does not take into account any effect from the
  thermal plasma, and gyrosynchrotron emission will in general deviate
  from a simple power law \cite[see, e.g.][]{fleishman10, kuznetsov11}. However, it serves to illustrate the feasibility of this mechanism in the case of I15103.

As in the previous hypothesis, the change in spectral index would
imply a change in the energy distribution of electrons, and all
spectral indices observed in I15103 are in principle
possible. Considering that the flux density has not changed much at 23
GHz between 2010 and 2013, equation \ref{eq_gyro} implies that
magnetic fields of several kG are necessary to flatten the radio
spectrum while maintaining a similar flux density at 23 GHz. These
magnetic field strengths have not been observed in PNe
\citep{leone14}, which makes this hypothesis highly unreliable.

\subsubsection{Opacity of free thermal electrons and Razin effect}

We have also explored the possibility that the change in the spectral index
were not due to a variation in the electron distribution, but to
selective absorption/suppression of radiation at low frequencies.  If
the emission on 2010-2011 is of gyrosynchrotron/synchrotron nature,
this can be later suppressed by an increasing amount of ionized gas
via two possible processes: a) opacity from free electrons or b) the Razin
(or Razin-Tsytovich) effect \citep{razin60, tsytovich51,
  ginzburg65}. Both processes are more efficient at lower frequencies. 

\begin{figure}
\epsscale{1}
\plotone{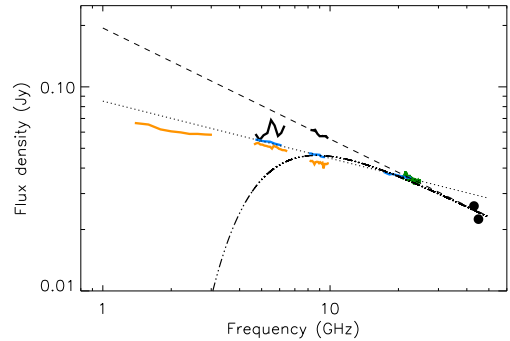}
\caption{Illustration of the effect of the opacity due to ionized gas. Dashed line: linear fit to the data on 2010-2011.  Dotted line: linear fit to the data on October 2012. 
Dashed-dotted line: Expected emission by applying the opacity of ionized gas ($\tau \propto \nu^{-2.1}$) to the fit to the 2010-2011 data. Color codes and symbols for the data are the same as in Fig.\ref{fig_observations}.
\label{fig_opacity} }
\end{figure}

In the former case (a), the opacity due to free electrons has a
dependency $\propto \nu^{-2.1}$. We tested this possibility against
our data, assuming a screen of ionized material between the observer
and the source of non-thermal emission, and applying an optical depth
$\tau_\nu = a\nu^{-2.1}$ to a SED with spectral index $-0.54$, varying
the constant $a$ until we reached a final spectral index of $\simeq
-0.28$ between 8 and 23 GHz. However, the resulting screen of ionized
gas would have such a high opacity at lower frequencies that we should
not detect any emission at $\simeq 1-2$ GHz
(Fig. \ref{fig_opacity}). Moreover, an optically thick screen of free
electrons would produce thermal free-free emission with a flux density
proportional to $\Omega\nu^2$, where $\Omega$ is the solid angle
subtended by the screen. Given the frequency dependence of free-free
emission, it cannot compensate the deficit of emission at 1-2 GHz
since, in that case, thermal emission would dominate at even higher
frequencies, with a positive spectral index that is not observed.

In the case of Razin effect (b), the shape of the spectral energy
distribution of synchrotron emission within a plasma takes the form

\begin{equation}
P(\epsilon,\beta) = \epsilon^{(\beta-1)/2}\int_0^\infty f
F\left(\frac{x}{f^3}\right) x^{(\beta-3)/2} dx
\label{razin_eq}
\end{equation}

\citep[function P in equation
\ref{razin_eq} in][but with some changes in the nomenclature of variables]{pacholczyk70}, where

\begin{equation}
f=\left(1+\frac{\epsilon}{x}\right)^{-1/2}
\end{equation}
\begin{equation}
\epsilon=\frac{20 n_e}{B \nu \sin\theta}.
\end{equation}

The function
$F$ is defined as

\begin{equation}
F(y) = y\int_y^\infty K_{5/3}(z) dz
\end{equation}

where $K_{5/3}$ is the Bessel function of order 5/3. 

The key parameter is the ratio $n_e/(B \sin\theta)$, which controls
the change in spectral slope. Assuming $\sin\theta\simeq 1$, the
critical frequency below which the synchrotron emission is mostly
suppressed is $\simeq 20 n_e/B$.  Applying Equation \ref{razin_eq} to
a SED with spectral index $-0.54$, to reach a final spectral index of
$\simeq -0.28$ between 8 and 23 GHz, would also imply the complete
suppression of emission at low frequency, as in the case of plasma
opacity (a), ruling out these two hypothesis.

\subsubsection{Combination of thermal and non-thermal emission}
\label{sec_razin}

A possible explanation for the observed change in spectral index is
the behavior predicted by \cite{fleishman06} for non-thermal emission
within dense plasmas when the magnetic field has both an ordered and a
random component. 

We note that the spectral index $\alpha\simeq
-0.28$ observed in 2012 is similar to that in the Crab nebula \citep{bietenholz97} and
other pulsar wind nebulae \citep{gaensler06}. In these cases, the
emission can be interpreted as ``jitter'' (or ``diffusive synchrotron'') 
radiation \citep{toptygin87,medvedev00,fleishman07}, 
i.e., electrons moving in a random magnetic field, as opposed to the
ordered field for usual synchrotron emission.

If we consider I15103 as having both an ordered and a random
component, where the ordered component dominates, relativistic
electrons moving along the magnetic field lines would mainly show a
normal synchrotron emission, with a relatively small contribution of
jitter emission. However, if this magnetic field is embedded in a
plasma, the Razin effect would suppress the synchrotron component. The
suppression of jitter emission depends much more slowly on plasma
density \citep[equation 40 in][]{fleishman06}, and therefore, if this
density is sufficiently high, the jitter emission will eventually
dominate the entire spectrum even though the magnetic field and the
electron energetics remained unchanged.  The Razin effect is actually
stronger in the presence of dense plasmas and relatively weak magnetic
fields, so we expect it to be more readily detectable under the
physical conditions of PNe, over those in other synchrotron-emitting
objects, such as radio galaxies or supernovae.

In this scenario, and for the case of I15103, until 2010, while the
photoionization of the PN was still low, it would show a mixture of
synchrotron and jitter emission (consistent with a magnetic field with
an ordered and a turbulent component). The increasing plasma density
of the newly formed ionized region of the PN would then suppress the
synchrotron component due to Razin effect. The final result would be
pure jitter emission in the 2012 data.  If the change in the spectral
slope is due to a sudden change in the dominant radio continuum
emission mechanism, it would suggest that we are witnessing the onset
of the photoionization in a PN. However, given that we do not have a direct
determination of the magnetic field strength, the plasma density, and
the size of the non-thermal emitting region, it is not possible at
this point to determine whether this interpretation is correct.


It is interesting to compare the flattening in the spectral energy
distribution of the radio continuum emission, with the flattening seen
in the post-AGB star IRAS 15445$-$5449
\citep{perezsanchez13_synchrotron}, whose nature is clearly different
from the one we see in I15103. In IRAS 15445$-$5449, the change in the
slope between 2.5 and 22 GHz is due to an increase of flux density at
high frequencies. This was interpreted as an increasingly higher
contribution of free-free emission which could arise from a
shock-ionized wind. It is known that a highly collimated ionized wind
can produce significant free-free emission \citep{reynolds86}, and a
stronger emission at high frequencies could be the consequence of a
sudden increase in its mass-loss rate. On the contrary, the flattening we
observe in I15103 is due to a decrease of flux density at low
frequencies. The flux density distribution seems to go down, pivoting
around $\nu\simeq 23$ GHz, where it does not change much. This cannot
be explained by an increase of free-free emission, but strongly
indicates a suppression of emission at low frequencies.
Other post-AGB stars show non-thermal emission \citep{bains09}, 
but there is no information on their time evolution, so a comparison 
is not possible at this stage.

Flat radio spectra with indices $\alpha > -0.5$ have also been found
in Wolf-Rayet stars \citep{chapman99}, and explained as a combination
of thermal and non-thermal emission in colliding wind binaries. Under
these models, the non-thermal component has $\alpha < -0.5$, and the
flattening would be attributed to the thermal mechanism. However, as mentioned
above, we cannot explain the progressive flattening of the spectrum in
I15103 with an increase of thermal emission.

\subsection{Processes taking place during the birth of a planetary nebula}

I15103 is the first known PN in which synchrotron emission could be
detected, since the free-free emission is so weak that it has not
masked the non-thermal emission yet. This
non-thermal emission due to shocks could share a common origin with
that seen in some post-AGB stars
\citep{bains09,perezsanchez13_synchrotron}.

However, in the case of I15103, if we assume the Coulomb losses as the
explanation for the variability in the spectral index (Sec. \ref{sec_coulomb}), we would be
observing the results of a jet launched close to the moment when the
post-AGB becomes a PN. Theoretical predictions affirm that jets can be
launched when magnetic field lines are twisted due to the different
rotation speeds of the envelope and the stellar core of an evolved
star, and the energy is released in the form of a collimated ejection
\citep{matt06}. The conditions for this magnetic explosion will be
favoured when a star enters the PN phase, as is the case of I15103,
since the magnetic field lines away from the star will tend to be
locked by the ionized gas to the slower rotational motions of the
outer envelope. Moreover, recent studies show that magnetic
  fields are stronger in older post-AGB stars and might be related to the origin
  of non-spherical morphologies in these objects \citep{gonidakis14}.

The changes in the spectral index could also be produced by the Razin
effect (Sec. \ref{sec_razin}), where the beginning of photoionization
could start new physical processes that affect the behavior of the
radio continuum emission.



Previous studies of radio continuum in very
young PNe showed that in several cases  \citep{cerrigone08, cerrigone11,
  casassus07} the observed spectral index could not be explained by
the classical optically thin free-free emission and the deviations
from the expected $\alpha$ were in some cases attributed to
calibration uncertainties or other instrumental effects. Radio continuum variability as
well as negative spectral indices were also found in young
PNe \citep{cerrigone11}, but limited time and frequency coverage
prevented a detailed identification of the processes taking place.  In
fact, it is possible that some of these young PN may be undergoing the same
processes as we are witnessing in I15103, but it has only been possible to
identify them in this study, thanks to the implementation of
new broad-band receivers and correlators in radio telescopes such as ATCA
or the Jansky Very Large Array.

\subsection{Predictions on source characteristics and their future evolution}

A monitoring of the radio spectrum of I15103 would shed light on
the processes taking place in this unique object.  As shown above,
possible origins of the variation seen in the spectral index are
changes in synchrotron emission due to Coulomb collisions
(Sec. \ref{sec_coulomb}), or selective suppression of part of the
emission due to Razin effect (Sec. \ref{sec_razin}). However, we
expect that these processes would lead to different future evolutions
in the short term. In the case of changes in the energy distribution
of electrons due to Coulomb collisions, we expect that the spectral
flattening will continue. Eventually, a low-frequency cutoff will
arise in the spectrum, reflecting a complete suppression of electrons
below an energy threshold.  On the other hand, if the physical process
is Razin suppression of synchrotron emission, leaving only jitter
emission, we do not expect a significant further flattening of the
emission, since the dominant radiative process after 2012 should not
be immediately affected by the Razin effect.

As seen in Fig. \ref{fig_map}, the possible contribution of free-free
emission from ionized material is still much weaker than the
non-thermal emission in this source. In the long term, we predict that
free-free radiation will significantly increase in the future, as
photoionization proceeds. Its more immediate effects would be an increase
of flux densities at high frequencies due to free-free emission, and a
decrease at low frequencies due to the absorption of non-thermal
emission by the plasma opacity. At later times, free-free emission
will eventually dominate the radio spectrum of the source, giving rise
to a thermal SED similar to other young PNe \citep{gomez05}, with no
trace of the non-thermal emission that we observe now.

\section{Conclusions}

We have observed radio continuum emission between 1 and 46 GHz towards
the young PN IRAS 15103$-$5754, in different epochs between 2010 and
2012, using the Australia Telescope Compact Array. Archive radio continuum data from
1991 to 2012 have been used to complete the information on this
source. Our main conclusions are as follow:
\begin{itemize}

\item Radio continuum emission shows a non-thermal spectrum, which is
  not typical of PNe.

\item The flux density seems to have decreased with time, especially at
  the lowest frequencies. In particular we see a flattening of the
  spectrum, with a spectral index $\alpha \simeq -0.54$ in 2010-2011,
  which changed to $\alpha\simeq -0.28$ in 2012. While the radio
  spectrum observed in 2010-2011 can be explained as arising from synchrotron
  emission with electrons accelerated in a shock, the one in 2012 is
  not compatible with this process alone.

\item The variation of the radio spectrum in I15103 can be
  attributed to two effects of the onset of photoionization in this source:

\begin{itemize}
\item The loss of energy of the synchrotron-emitting
  electrons due to Coulomb collisions within the ionized region of the
  PN. The estimated lifetime of the electrons is 18 yr, consistent
  with the timescale of variation of the flux density in the archival
  data. The onset of the photoionization could produce the sudden
  change visible in the spectral index between 2010 and 2012.

\item The electrons moving in a magnetic field with both an ordered
  and a random component, which would give rise to a mixture of
  synchrotron and "jitter" emission. The Razin effect would suppress
  the synchrotron component as photoionization progresses. Only the jitter
  component with a flatter spectral index, will remain.

\end{itemize}


\item We postulate that the flattening of non-thermal spectra could be
  a signature of the onset of the PN phase.
\end{itemize}


\acknowledgments

The Australia Telescope Compact
Array is part of the Australia Telescope National Facility which is
funded by the Commonwealth of Australia for operation as a National
Facility managed by CSIRO. This paper includes archived data obtained
through the Australia Telescope Online Archive
(http://atoa.atnf.csiro.au). 

The authors thank 
Antonio Alberdi, Olivier Chesnau, Gregory Fleishman,
Jose L. G\'omez,  Athanasios Katsiyannis, and Miguel A. P\'erez-Torres
 for fruitful discussions. They also thank James Urquhart and
Tara Murphy for providing images from RMS and MGPS2 surveys. JFG
wishes to express his gratitude to CSIRO Astronomy and Space Science,
and the Observatoire de la C\^ote d'Azur for their support and
hospitality during the preparation of this paper. 

The authors acknowledge financial support from MICINN (Spain) grants
AYA2011-29754-C03-02, AYA2011-30228-C03 (both including FEDER funds),
CSD2009-00038, AYA2009-07304, AYA2012-32032, and AYA2014-57369-C3-3-P,
from grant PE9-1160 of the Greek General Secretariat for Research and
Technology, from CONACyT and PROMEP (Mexico), from Ministerio de
Educaci\'on (Spain) under Programa Nacional de Movilidad de Recursos
Humanos del Plan Nacional de I+D+I 2008-2011, and from grant 12VI20 of
the Universidad de Vigo.

{\it Facilities:}\facility{ATCA}

\begin{thebibliography}{60}
\expandafter\ifx\csname natexlab\endcsname\relax\def\natexlab#1{#1}\fi

\bibitem[{{Achterberg}(2000)}]{achterberg00}
{Achterberg}, A. 2000, in IAU Symposium, Vol. 195, Highly Energetic Physical
  Processes and Mechanisms for Emission from Astrophysical Plasmas, ed.
  P.~C.~H. {Martens}, S.~{Tsuruta}, \& M.~A. {Weber}, 291

\bibitem[{{Bains} {et~al.}(2009){Bains}, {Cohen}, {Chapman}, {Deacon}, \&
  {Redman}}]{bains09}
{Bains}, I., {Cohen}, M., {Chapman}, J.~M., {Deacon}, R.~M., \& {Redman}, M.~P.
  2009, \mnras, 397, 1386

\bibitem[{{Bains} {et~al.}(2003){Bains}, {Gledhill}, {Yates}, \&
  {Richards}}]{bains03}
{Bains}, I., {Gledhill}, T.~M., {Yates}, J.~A., \& {Richards}, A.~M.~S. 2003,
  \mnras, 338, 287

\bibitem[{{Bains} {et~al.}(2004){Bains}, {Richards}, {Gledhill}, \&
  {Yates}}]{bains04}
{Bains}, I., {Richards}, A.~M.~S., {Gledhill}, T.~M., \& {Yates}, J.~A. 2004,
  \mnras, 354, 529

\bibitem[{{Bendjoya} {et~al.}(2014){Bendjoya}, {Su{\'a}rez}, {G{\'o}mez},
  {Guerrero}, {Miranda}, {Green}, {Uscanga}, {Rizzo}, \&
  {Ramos-Larios}}]{bendjoya14}
{Bendjoya}, P., {Su{\'a}rez}, O., {G{\'o}mez}, J.~F., {et~al.} 2014, in
  Asymmetrical Planetary Nebulae VI conference, Proceedings of the conference
  held 4-8 November, 2013. Edited by C. Morisset, G. Delgado-Inglada and S.
  Torres-Peimbert. http://www.astroscu.unam.mx/apn6/PROCEEDINGS/, 5

\bibitem[{{Bietenholz} {et~al.}(1997){Bietenholz}, {Kassim}, {Frail}, {Perley},
  {Erickson}, \& {Hajian}}]{bietenholz97}
{Bietenholz}, M.~F., {Kassim}, N., {Frail}, D.~A., {et~al.} 1997, \apj, 490,
  291

\bibitem[{{Bobrowsky} {et~al.}(1998){Bobrowsky}, {Sahu}, {Parthasarathy}, \&
  {Garc\'\i a-Lario}}]{bobrowsky98}
{Bobrowsky}, M., {Sahu}, K.~C., {Parthasarathy}, M., \& {Garc\'\i a-Lario}, P.
  1998, \nat, 392, 469

\bibitem[{{Bridle} \& {Perley}(1984)}]{bridle84}
{Bridle}, A.~H., \& {Perley}, R.~A. 1984, \araa, 22, 319

\bibitem[{{Carrasco-Gonz{\'a}lez} {et~al.}(2010){Carrasco-Gonz{\'a}lez},
  {Rodr{\'{\i}}guez}, {Anglada}, {Mart{\'{\i}}}, {Torrelles}, \&
  {Osorio}}]{carrascogonzalez10}
{Carrasco-Gonz{\'a}lez}, C., {Rodr{\'{\i}}guez}, L.~F., {Anglada}, G., {et~al.}
  2010, Science, 330, 1209

\bibitem[{{Casassus} {et~al.}(2007){Casassus}, {Nyman}, {Dickinson}, \&
  {Pearson}}]{casassus07}
{Casassus}, S., {Nyman}, L.-{\AA}., {Dickinson}, C., \& {Pearson}, T.~J. 2007,
  \mnras, 382, 1607

\bibitem[{{Cerrigone} {et~al.}(2011){Cerrigone}, {Trigilio}, {Umana}, {Buemi},
  \& {Leto}}]{cerrigone11}
{Cerrigone}, L., {Trigilio}, C., {Umana}, G., {Buemi}, C.~S., \& {Leto}, P.
  2011, \mnras, 412, 1137

\bibitem[{{Cerrigone} {et~al.}(2008){Cerrigone}, {Umana}, {Trigilio}, {Leto},
  {Buemi}, \& {Hora}}]{cerrigone08}
{Cerrigone}, L., {Umana}, G., {Trigilio}, C., {et~al.} 2008, \mnras, 390, 363

\bibitem[{{Chapman} {et~al.}(1999){Chapman}, {Leitherer}, {Koribalski},
  {Bouter}, \& {Storey}}]{chapman99}
{Chapman}, J.~M., {Leitherer}, C., {Koribalski}, B., {Bouter}, R., \& {Storey},
  M. 1999, \apj, 518, 890

\bibitem[{{Cohen} {et~al.}(2006){Cohen}, {Chapman}, {Deacon}, {Sault},
  {Parker}, \& {Green}}]{cohen06}
{Cohen}, M., {Chapman}, J.~M., {Deacon}, R.~M., {et~al.} 2006, \mnras, 369, 189

\bibitem[{{de Marco}(2009)}]{demarco09}
{de Marco}, O. 2009, \pasp, 121, 316

\bibitem[{{Dgani} \& {Soker}(1998)}]{dgani98}
{Dgani}, R., \& {Soker}, N. 1998, \apjl, 499, L83

\bibitem[{{Drury}(1983)}]{drury83}
{Drury}, L.~O. 1983, Reports on Progress in Physics, 46, 973

\bibitem[{{Dulk}(1985)}]{dulk85}
{Dulk}, G.~A. 1985, \araa, 23, 169

\bibitem[{{Fleishman}(2006)}]{fleishman06}
{Fleishman}, G.~D. 2006, in Lecture Notes in Physics, Berlin Springer Verlag,
  Vol. 687, Geospace Electromagnetic Waves and Radiation, ed. J.~W. {Labelle}
  \& R.~A. {Treumann}, 87--99

\bibitem[{{Fleishman} \& {Bietenholz}(2007)}]{fleishman07}
{Fleishman}, G.~D., \& {Bietenholz}, M.~F. 2007, \mnras, 376, 625

\bibitem[{{Fleishman} \& {Kuznetsov}(2010)}]{fleishman10}
{Fleishman}, G.~D., \& {Kuznetsov}, A.~A. 2010, \apj, 721, 1127

\bibitem[{{Gaensler} \& {Slane}(2006)}]{gaensler06}
{Gaensler}, B.~M., \& {Slane}, P.~O. 2006, \araa, 44, 17

\bibitem[{{Gaensler} {et~al.}(2000){Gaensler}, {Stappers}, {Frail}, {Moffett},
  {Johnston}, \& {Chatterjee}}]{gaensler00}
{Gaensler}, B.~M., {Stappers}, B.~W., {Frail}, D.~A., {et~al.} 2000, \mnras,
  318, 58

\bibitem[{{Garc{\'{\i}}a-Segura} {et~al.}(2005){Garc{\'{\i}}a-Segura},
  {L{\'o}pez}, \& {Franco}}]{garciasegura05}
{Garc{\'{\i}}a-Segura}, G., {L{\'o}pez}, J.~A., \& {Franco}, J. 2005, \apj,
  618, 919

\bibitem[{{Ginzburg} \& {Syrovatskii}(1965)}]{ginzburg65}
{Ginzburg}, V.~L., \& {Syrovatskii}, S.~I. 1965, \araa, 3, 297

\bibitem[{{G{\'o}mez} {et~al.}(2005){G{\'o}mez}, {de Gregorio-Monsalvo},
  {Lovell}, {Anglada}, {Miranda}, {Su{\'a}rez}, {Torrelles}, \&
  {G{\'o}mez}}]{gomez05}
{G{\'o}mez}, J.~F., {de Gregorio-Monsalvo}, I., {Lovell}, J.~E.~J., {et~al.}
  2005, \mnras, 364, 738

\bibitem[{{G{\'o}mez} {et~al.}(2015){G{\'o}mez}, {Su{\'a}rez}, {Bendjoya},
  {Rizzo}, {Miranda}, {Green}, {Uscanga}, {Garc{\'{\i}}a-Garc{\'{\i}}a},
  {Lagadec}, {Guerrero}, \& {Ramos-Larios}}]{gomez15}
{G{\'o}mez}, J.~F., {Su{\'a}rez}, O., {Bendjoya}, P., {et~al.} 2015, \apj, 799,
  186

\bibitem[{{Gonidakis} {et~al.}(2014){Gonidakis}, {Chapman}, {Deacon}, \&
  {Green}}]{gonidakis14}
{Gonidakis}, I., {Chapman}, J.~M., {Deacon}, R.~M., \& {Green}, A.~J. 2014,
  \mnras, 443, 3819

\bibitem[{{Green} {et~al.}(2012){Green}, {McClure-Griffiths}, {Caswell},
  {Robishaw}, \& {Harvey-Smith}}]{green12}
{Green}, J.~A., {McClure-Griffiths}, N.~M., {Caswell}, J.~L., {Robishaw}, T.,
  \& {Harvey-Smith}, L. 2012, \mnras, 425, 2530

\bibitem[{{Gunn} {et~al.}(1994){Gunn}, {Spencer}, {Abdul Aziz}, {Doyle},
  {Davis}, \& {Pavelin}}]{gunn94}
{Gunn}, A.~G., {Spencer}, R.~E., {Abdul Aziz}, H., {et~al.} 1994, \aap, 291,
  847

\bibitem[{{Imai} {et~al.}(2007){Imai}, {Sahai}, \& {Morris}}]{imai07}
{Imai}, H., {Sahai}, R., \& {Morris}, M. 2007, \apj, 669, 424

\bibitem[{{Kuznetsov} {et~al.}(2011){Kuznetsov}, {Nita}, \&
  {Fleishman}}]{kuznetsov11}
{Kuznetsov}, A.~A., {Nita}, G.~M., \& {Fleishman}, G.~D. 2011, \apj, 742, 87

\bibitem[{{Kwok} {et~al.}(1981){Kwok}, {Purton}, \& {Keenan}}]{kwok81}
{Kwok}, S., {Purton}, C.~R., \& {Keenan}, D.~W. 1981, \apj, 250, 232

\bibitem[{{Lagadec} {et~al.}(2011){Lagadec}, {Verhoelst}, {M{\'e}karnia},
  {Su{\'a}eez}, {Zijlstra}, {Bendjoya}, {Szczerba}, {Chesneau}, {van Winckel},
  {Barlow}, {Matsuura}, {Bowey}, {Lorenz-Martins}, \& {Gledhill}}]{lagadec11}
{Lagadec}, E., {Verhoelst}, T., {M{\'e}karnia}, D., {et~al.} 2011, \mnras, 417,
  32

\bibitem[{{Lee} \& {Gary}(2000)}]{lee00}
{Lee}, J., \& {Gary}, D.~E. 2000, \apj, 543, 457

\bibitem[{{Leone} {et~al.}(2014){Leone}, {Corradi}, {Mart{\'{\i}}nez
  Gonz{\'a}lez}, {Asensio Ramos}, \& {Manso Sainz}}]{leone14}
{Leone}, F., {Corradi}, R.~L.~M., {Mart{\'{\i}}nez Gonz{\'a}lez}, M.~J.,
  {Asensio Ramos}, A., \& {Manso Sainz}, R. 2014, \aap, 563, A43

\bibitem[{{Lumsden} {et~al.}(2013){Lumsden}, {Hoare}, {Urquhart}, {Oudmaijer},
  {Davies}, {Mottram}, {Cooper}, \& {Moore}}]{lumsden13}
{Lumsden}, S.~L., {Hoare}, M.~G., {Urquhart}, J.~S., {et~al.} 2013, \apjs, 208,
  11

\bibitem[{{Matt} {et~al.}(2006){Matt}, {Frank}, \& {Blackman}}]{matt06}
{Matt}, S., {Frank}, A., \& {Blackman}, E.~G. 2006, \apjl, 647, L45

\bibitem[{{Medvedev}(2000)}]{medvedev00}
{Medvedev}, M.~V. 2000, \apj, 540, 704

\bibitem[{{Melnikov} \& {Magun}(1998)}]{melnikov98}
{Melnikov}, V.~F., \& {Magun}, A. 1998, \solphys, 178, 153

\bibitem[{{Mirabel} {et~al.}(1992){Mirabel}, {Rodriguez}, {Cordier}, {Paul}, \&
  {Lebrun}}]{mirabel92}
{Mirabel}, I.~F., {Rodriguez}, L.~F., {Cordier}, B., {Paul}, J., \& {Lebrun},
  F. 1992, \nat, 358, 215

\bibitem[{{Miranda} {et~al.}(2001){Miranda}, {G{\'o}mez}, {Anglada}, \&
  {Torrelles}}]{miranda01}
{Miranda}, L.~F., {G{\'o}mez}, Y., {Anglada}, G., \& {Torrelles}, J.~M. 2001,
  \nat, 414, 284

\bibitem[{{Murphy} {et~al.}(2007){Murphy}, {Mauch}, {Green}, {Hunstead},
  {Piestrzynska}, {Kels}, \& {Sztajer}}]{murphy07}
{Murphy}, T., {Mauch}, T., {Green}, A., {et~al.} 2007, \mnras, 382, 382

\bibitem[{{Nordhaus} \& {Blackman}(2006)}]{nordhaus06}
{Nordhaus}, J., \& {Blackman}, E.~G. 2006, \mnras, 370, 2004

\bibitem[{{Pacholczyk}(1970)}]{pacholczyk70}
{Pacholczyk}, A.~G. 1970, {Radio astrophysics. Nonthermal processes in galactic
  and extragalactic sources}

\bibitem[{{P{\'e}rez-S{\'a}nchez} {et~al.}(2013){P{\'e}rez-S{\'a}nchez},
  {Vlemmings}, {Tafoya}, \& {Chapman}}]{perezsanchez13_synchrotron}
{P{\'e}rez-S{\'a}nchez}, A.~F., {Vlemmings}, W.~H.~T., {Tafoya}, D., \&
  {Chapman}, J.~M. 2013, \mnras, 436, L79

\bibitem[{{Pestalozzi} {et~al.}(2000){Pestalozzi}, {Benz}, {Conway}, \&
  {G{\"u}del}}]{pestalozzi00}
{Pestalozzi}, M.~R., {Benz}, A.~O., {Conway}, J.~E., \& {G{\"u}del}, M. 2000,
  \aap, 353, 569

\bibitem[{{Razin}(1960)}]{razin60}
{Razin}, V.~A. 1960, Iz. Vys. Ucheb. Zaved. Radiofiz., 584, 3

\bibitem[{{Rephaeli}(1979)}]{rephaeli79}
{Rephaeli}, Y. 1979, \apj, 227, 364

\bibitem[{{Reynolds}(1986)}]{reynolds86}
{Reynolds}, S.~P. 1986, \apj, 304, 713

\bibitem[{{Sabin} {et~al.}(2007){Sabin}, {Zijlstra}, \& {Greaves}}]{sabin07}
{Sabin}, L., {Zijlstra}, A.~A., \& {Greaves}, J.~S. 2007, \mnras, 376, 378

\bibitem[{{Sahai} \& {Trauger}(1998)}]{sahai98}
{Sahai}, R., \& {Trauger}, J.~T. 1998, \aj, 116, 1357

\bibitem[{{Sarazin}(1999)}]{sarazin99}
{Sarazin}, C.~L. 1999, \apj, 520, 529

\bibitem[{{Shu}(1991)}]{shu91}
{Shu}, F.~H. 1991, {Physics of Astrophysics, Vol. I} (University Science Books)

\bibitem[{{Soker} \& {Rappaport}(2000)}]{soker00}
{Soker}, N., \& {Rappaport}, S. 2000, \apj, 538, 241

\bibitem[{{Tafoya} {et~al.}(2009){Tafoya}, {G{\'o}mez}, {Patel}, {Torrelles},
  {G{\'o}mez}, {Anglada}, {Miranda}, \& {de Gregorio-Monsalvo}}]{tafoya09}
{Tafoya}, D., {G{\'o}mez}, Y., {Patel}, N.~A., {et~al.} 2009, \apj, 691, 611

\bibitem[{{Toptygin} \& {Fleishman}(1987)}]{toptygin87}
{Toptygin}, I.~N., \& {Fleishman}, G.~D. 1987, \apss, 132, 213

\bibitem[{{Tsytovich}(1951)}]{tsytovich51}
{Tsytovich}, V.~N. 1951, Vestn. Mosk. Univ., 11, 27

\bibitem[{{Urquhart} {et~al.}(2007){Urquhart}, {Busfield}, {Hoare}, {Lumsden},
  {Clarke}, {Moore}, {Mottram}, \& {Oudmaijer}}]{urquhart07}
{Urquhart}, J.~S., {Busfield}, A.~L., {Hoare}, M.~G., {et~al.} 2007, \aap, 461,
  11

\bibitem[{{van de Steene} \& {Pottasch}(1993)}]{vdsteene93}
{van de Steene}, G.~C.~M., \& {Pottasch}, S.~R. 1993, \aap, 274, 895

\end{thebibliography}

  \end{document}